\documentclass[12pt]{iopart}
\usepackage{graphicx}
\usepackage{hyperref}
\usepackage{bigstrut}
\usepackage{multirow}
\usepackage{iopams}  
\expandafter\let\csname equation*\endcsname\relax
\expandafter\let\csname endequation*\endcsname\relax
\usepackage{amsmath}

\begin{document}

\title[Quantum state revivals in quantum walks on cycles]{Quantum state revivals in quantum walks on cycles}

\author{Phillip R. Dukes}

\address{University of Texas at Brownsville, Brownsville, TX 78520, USA}
\address{4 November 2014}
\ead{phil.dukes@utb.edu}

\begin{abstract}
Recurrence in the classical random walk is well known and described by the P\'{o}lya number. For quantum walks, recurrence is similarly understood in terms of the probability of a localized quantum walker to return to its origin. Under certain circumstances the quantum walker may also return to an arbitrary initial quantum state in a finite number of steps. Quantum state revivals in quantum walks on cycles using coin operators which are constant in time and uniform across the path have been described before but only incompletely. In this paper we find the general conditions for which full-quantum state revival will occur. 
\end{abstract}

\section{Introduction}
A quantum walk is the quantum-mechanical complement to the classical random walk. In a quantum walk the ``walker'' evolves according to a unitary transformation between initial and final states, either in discrete steps of time or by a continuous-time evolution under a Hamiltonian operator. The discrete-time quantum walk was first described by Aharonov et al. \cite{Aharonov:1} where it was noted that due to quantum interference effects the average path length of the quantum walk can be longer than the maximum allowed path length in a classical random walk. To take advantage of this phenomenon the quantum walk has since been applied to the development of quantum search algorithms \cite{Ambainis:2, Kendon:1, Ambainis:1, Venegas:1} in terms of both the discrete-time \cite{Shenvi:1, Kempe:1} and continuous-time \cite{Farhi:1, Childs:1} quantum walks. Both the discrete and continuous quantum walks have also been shown to be universal for quantum computation \cite{Childs:2, Lovett:1, Underwood:1}.

An important problem in the study of classical random walks is determining the probability of the walker returning to its origin.  This is referred to as recurrence and is determined by the random walk's P\'{o}lya number \cite{Polya:1, Finch:1, Novac:1}. Recurrence in a quantum walk is similarly defined as the probability after $N$ steps for observing the quantum walker at its point of  origin \cite{Chandrashekar:1, Chou:1, Segawa:1}. Recurrence in continuous-time quantum walks has also been studied \cite{Darazs:1}.

The criterion of the quantum walker returning to its initial quantum state or quantum state revival is a more stringent requirement. Previous work has looked at full revivals in quantum walks in a 2 dimensional graph \cite{Stefanak:1}. A similar problem looking at quantum diffusion on a cyclic lattice is also treated \cite{Goyeneche:1}. This paper is concerned with the conditions under which a quantum walker in an arbitrary quantum state on a $k$-cycle with $k$ unitary transformation sites, will return to its initial quantum state in $N$ steps. Our assumptions are that each of the unitary transformations be time independent and equal. Quantum state revivals occur when the $k$-cycle operator, $U_k\left( \rho,\alpha,\beta\right)$ satisfies  $U_{k}^{N} = I_{2k}$ where $I_{2k}$ is the $2k \times 2k$ identity matrix.

{\section{Discrete quantum walks in one dimension}
The necessary elements of the classical random walk are a walker and a random coin toss mechanism. For each toss of the coin the walker takes a step to the right if ``heads'' or a step to the left if ``tails''. An important distinction of the quantum walk is the quantum property of superposition, in this case a superposition of the amplitudes corresponding to a step to the left and a step to the right. Thus the quantum counterpart to the classical random walk involves a quantum walker with a two state coin space and a unitary coin operator. The coin operator can be continuously tuned in both how much it rotates the original state and its relative phase change. The essential quantum behavior is typically modeled in terms of a quantum two state system such as a spin $\frac{1}{2}$ particle for the walker and a general $2 \times 2$ unitary transformation matrix for the coin operator. It is possible that the coin operator may change with time or have different coin operators at each discrete position of the walk \cite{Brun:1, Shikano:1, Tregenna:1}. In all that follows, however, we will only consider a coin operator which is constant in time and uniform for all positions.

\subsection{Discrete quantum walks on a line}
For concreteness, we consider a two state quantum walker located at the origin of a line extending in the positive and negative directions. Each step of the walker has equal length and occurs at discrete time intervals. Let $\textbf{H}z$  represent the Hilbert space of the locations of the walker along the infinite line. This space is spanned by the basis states 
$\lbrace\vert i \rangle : i \in \mathbb{Z}\rbrace $, such that $\vert i \rangle $ corresponds with a walker localized at position $i$ on the line. The coin space $\textbf{H}c$ of our quantum walker will be spanned by basis states $\lbrace\vert \uparrow \rangle, \vert \downarrow \rangle\rbrace$. The Hilbert space for the walker system will now be $\textbf{H}w =  \textbf{H}z \otimes \textbf{H}c$, the tensor product of the position space with the coin space. In our model the spin-up and spin-down amplitudes will step in opposite directions along the line such that
\begin{subequations}
\begin{align}
\vert i, \uparrow \rangle &\longrightarrow \vert i-1, \uparrow \rangle \label{eq:tensorstep1}\\
\vert i, \downarrow \rangle &\longrightarrow \vert i+1 , \downarrow \rangle 
 \label{eq:tensorstep2}
 \end{align}
\end{subequations}
a spin-up amplitude steps in the negative direction (to the left) and a spin-down amplitude steps in the positive direction (to the right). The conditional shift operator in $\textbf{H}w$ which does this is 
\begin{subequations}
\begin{align}
S_{Z} &=  \displaystyle\sum_{i=-\infty}^{\infty} \vert i-1 \rangle \langle i \vert \otimes \vert \uparrow \rangle \langle \uparrow \vert  + \displaystyle\sum_{i=-\infty}^{\infty} \vert i+1 \rangle \langle i \vert \otimes \vert \downarrow \rangle \langle \downarrow \vert \label{eq:Szop}\\
 & = \displaystyle \sum_{s=0}^{1} \displaystyle \sum_{i=-\infty}^{\infty}\vert i+2s-1 \rangle \langle i \vert \otimes  \vert s \rangle \langle s \vert,
\label{eq:Szopshort}
 \end{align}
\end{subequations}
expression \ref{eq:Szopshort} is expressed in the quantum computational basis for which $ \vert 0 \rangle=\vert \uparrow \rangle $ and  $ \vert 1 \rangle =  \vert \downarrow \rangle $.

Prior to taking each step, the coin operator would be applied to the walker's amplitude at each position $ \vert i \rangle$ effectively rotating the spin-state into a coherent superposition of spin-up and spin-down amplitudes and thus control the portion of amplitude which is shifted to the left and to the right. A parameterization of the most general $2 \times 2$ unitary coin operator, to within a global phase change, is \cite{Tregenna:1}
\begin{equation}
C_{2} = \left(  \begin{array}{c c}
\sqrt{\rho} & \sqrt{1-\rho}\, e^{i\alpha} \\
\sqrt{1-\rho}\, e^{i\beta} & -\sqrt{\rho}\, e^{i\left(  \alpha + \beta\right)}
\end{array}\right),\; 0\leq\rho\leq1,\; 0\leq\alpha, \beta\leq\pi.
\label{eq:C2op}
\end{equation}
The complete operator for each step of the discrete quantum walk on the infinite line is then
\begin{equation}
U_{Z} = S_{Z} \cdot \left( I_{Z} \otimes C_{2} \right).
\label{eq:Uzop}
\end{equation}
This provides a uniform application of the coin operator $C_{2}$ across all the possible positions of the walker.

An often cited coin operator is the $2 \times 2$ Hadamard operator  \cite{Venegas:1, Lovett:1, Kempe:2}
\begin{equation}
C_{2}\left(\rho = \frac{1}{2},\alpha = 0,\beta =0\right) =\frac{1}{\sqrt{2}} \left(
\begin{array}{c c}
1 & 1 \\
1 & -1
\end{array}\right).
\label{eq:hadaop}
\end{equation}
Fig. \ref{fig:QWonLine} illustrates the four step evolution of a quantum walk on a line with the initial state  $\vert \psi_{i}\rangle = \vert 0, \uparrow \rangle $. Each step of the quantum walk is divided into a unitary transformation using the Hadamard coin operator of Eq. (\ref{eq:hadaop}) immediately followed by the conditional shift operator in Eq. (\ref{eq:Szopshort}).

\begin{figure}[h!]
\includegraphics[scale=0.36]{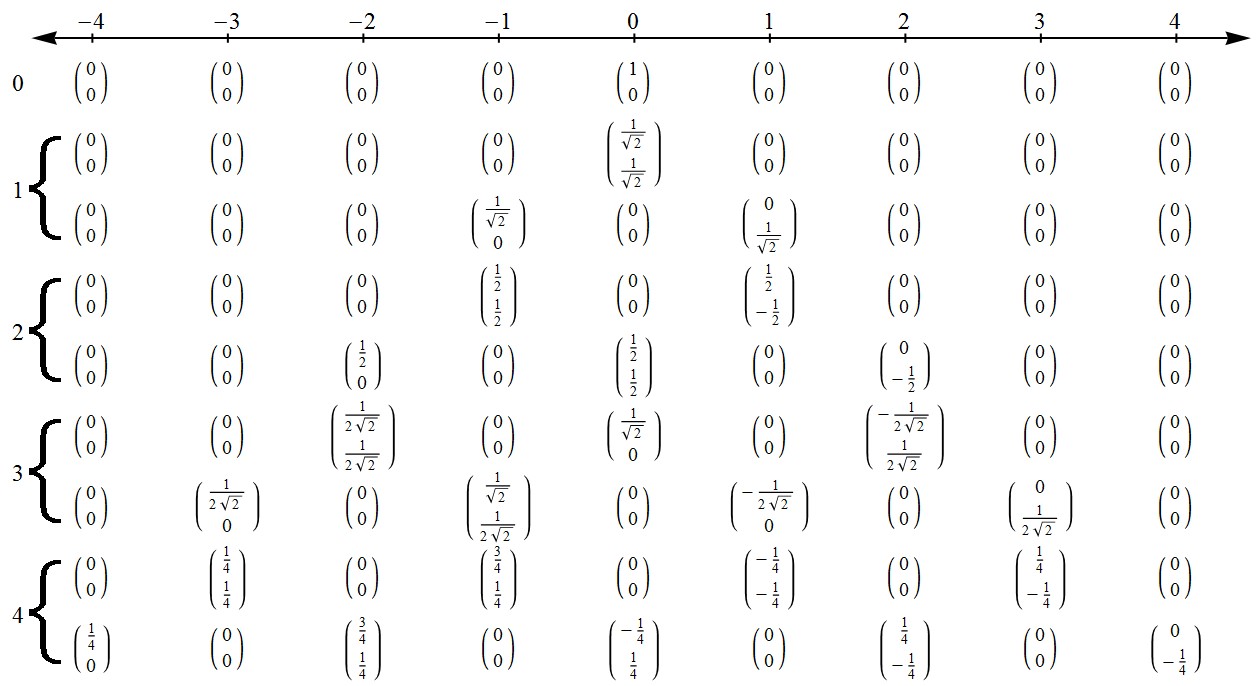}
\caption{The first four steps of a quantum walk on an infinite line. The initial state $\vert \psi_{i}\rangle =\vert 0, \uparrow \rangle $. Each step of the quantum walk consists of the unitary transformation of a $2 \times 2$ Hadamard coin operator followed by the conditional shift operator.}
\label{fig:QWonLine}
\end{figure}
As the walk progresses, an asymmetry in the amplitudes skewing the probabilities for finding the walker at locations on the left side of the initial position becomes evident on completion of the third step. The probabilities will be skewed to the right with an initial state of $\vert 0, \downarrow \rangle $. The asymmetry arises from the fact that the Hadamard operator treats the two states $\vert \uparrow \rangle$ and $\vert \downarrow \rangle$ differently by inducing a phase inversion in the $\vert \downarrow \rangle$ amplitude. The Hadamard operator will develop a symmetric walk in the probabilities with $\vert \psi_{i}\rangle =\left( \frac{1}{\sqrt{2}}\vert 0, \uparrow \rangle + \frac{i}{\sqrt{2}}\vert 0, \downarrow \rangle \right) $.

\subsection{Discrete quantum walks on a cycle}
The conditional shift operator in Eq. (\ref{eq:Szopshort}) can be readily modified to operate on a cycle or closed loop of $k$ steps as
\begin{equation}
S_{k} =\displaystyle\sum_{s=0}^{1}\displaystyle\sum_{i=0}^{k-1} \vert i +2s -1 \pmod{k} \rangle \langle i \vert \otimes \vert s \rangle \langle s \vert
\label{eq:Skop}
\end{equation}

\begin{equation}
U_{k} = S_{k} \cdot \left( I_{k} \otimes C_{2}  \right),
\label{eq:Ukop}
\end{equation}
where $I_{k}$ is the $k \times k$ identity matrix. $U_{k}$ can be expressed as a $(2k) \times (2k)$ matrix. Consider the $k=3$ loop, the operator in Eq. (\ref{eq:Ukop}) becomes,

\footnotesize
\begin{equation}
U_{3} =  \begin{pmatrix}
0 & 0 & \sqrt{\rho} & \sqrt{1-\rho}\,e^{i\alpha} & 0 & 0\\
0 & 0 & 0 & 0 & \sqrt{1-\rho}\,e^{i\beta} & -\sqrt{\rho}\,e^{i\left(\alpha+\beta \right)}\\
0 & 0 & 0 & 0 & \sqrt{\rho} & \sqrt{1-\rho}\,e^{i\alpha}\\
\sqrt{1-\rho}\,e^{i\beta} & -\sqrt{\rho}\,e^{i\left(\alpha+\beta \right)} & 0 & 0 & 0 & 0\\
\sqrt{\rho} & \sqrt{1-\rho}\,e^{i\alpha} & 0 & 0 & 0 & 0\\
0 & 0 & \sqrt{1-\rho}\,e^{i\beta} & -\sqrt{\rho}\,e^{i\left(\alpha+\beta \right)} & 0 & 0 
\end{pmatrix}
\label{eq:Uk=3}
.\end{equation}
\normalsize

Fig. \ref{fig:QWonk=3} shows a quantum walk on a $k=3$ cycle using the coin operator $ C_{2}\left(\rho=\frac{2}{3},\alpha=0,\beta=0 \right) $.
\begin{figure}[h!]
\centering
\includegraphics[scale=0.6]{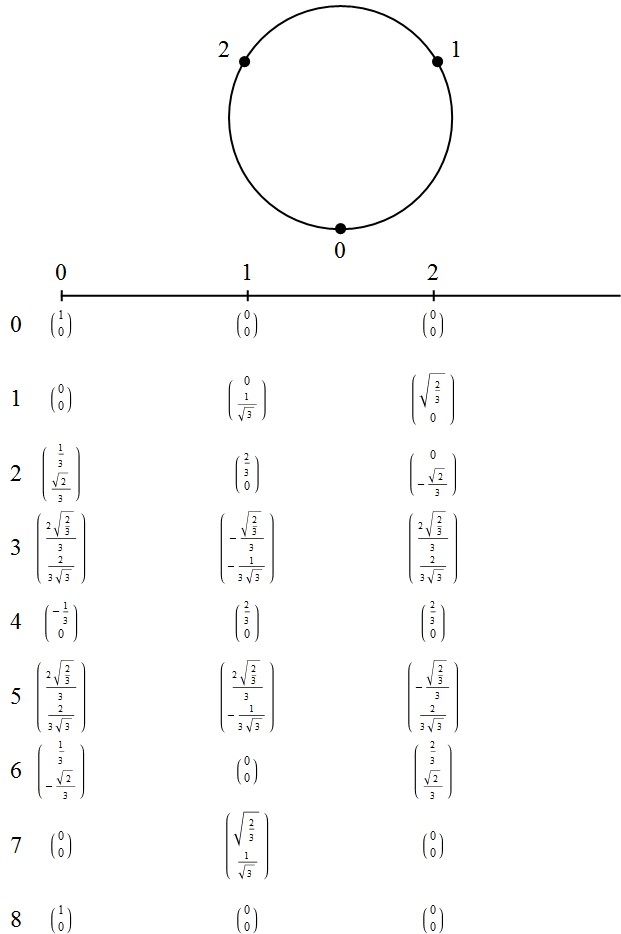}
\caption{The first eight steps of a quantum walk on a $k=3$ cycle. The initial state is  $\vert \psi_{i}\rangle =\vert 0, \uparrow \rangle $. The quantum walk operator is $ U_{3}\left(\rho=\frac{2}{3},\alpha=0,\beta=0 \right) $.}
\label{fig:QWonk=3}
\end{figure}

\section{Conditions for quantum state revivals}
In Fig. \ref{fig:QWonk=3} we see it is possible, with certain choices of a constant and uniform coin operator, for the quantum walk on a cycle to return to its initial quantum state within a finite number of steps. The occurrence of quantum state revivals in quantum walks on cycles was probably first mentioned in the literature by Travaglione and Milburn \cite{Travaglione:1} where they noted a revival in eight steps on a cycle with $k=4$. Later, Tregenna et.al. \cite{Tregenna:1} found a handful of other instances. In this paper we wish to establish the general conditions for quantum state revivals in quantum walks on cycles. We observe that the operator $U_{k}$ is a $2 \times 2$ block-circulant matrix \cite{Wyn-jones:1, Davis:1} and is the generator of a unitary cyclic group. When the unitary operator $U_{k}\left( \rho, \alpha, \beta\right) $ generates a finite cyclic group, quantum state revival will occur.  

\subsection{Circulant matrices}
\begin{quote}
If you run into a circulant in the course of a problem you are happy to make its acquaintance. --- Persi Diaconis
\end{quote}
Of the many interesting properties circulant matrices have the following are the most important for us:\cite{Diaconis:1}
\begin{itemize}
  \item The class circulant is closed under product, transpose, and inverse operations.
  \item All circulants are simultaneously  diagonalized by the Fourier matrix.
\end{itemize}

Due to its circulant symmetries only a single row or column of a circulant matrix is required to specify it. The first row or column of a circulant matrix is referred to as its circulant vector $\nu$. For an $M \times M$ circulant matrix $A = \left( a_{m,n}\right)$ the first row circulant vector is $\nu = \left( a_{0},a_{1}, \ldots, a_{M-1} \right)$ where we denote the first position with $0$ as a matter of convenience. The circulant matrix $A$ can then be represented as

\begin{equation}
A = \left( a_{\left( n-m\right) \left(\bmod{M}\right)  } \right)_{m,n} = CIRC_{M}\left(a_{0},a_{1}, \ldots, a_{M-1} \right).
\label{eq:CIRC(A)}
\end{equation}

The quantum walk operator $U_{k}$ of Eq. (\ref{eq:Ukop}) is $2 \times 2$ block circulant and is represented in terms of a circulant vector of $2 \times 2$ matrices,
\footnotesize
\begin{equation}
U_{k} = \mathrm{CIRC}_{k}\left( \begin{bmatrix}0 & 0\\0 &0\end{bmatrix}_{0}, \begin{bmatrix}\sqrt{\rho} & \sqrt{1-\rho}\,e^{i \alpha}\\0 &0\end{bmatrix}_{1}, \begin{bmatrix}0 & 0\\0 &0\end{bmatrix}_{2}, \:\cdot\,\cdot\,\cdot\: , \begin{bmatrix}0 & 0\\\sqrt{1-\rho}\, e^{i \beta} &-\sqrt{\rho}\,e^{i \left( \alpha + \beta \right) }\end{bmatrix}_{k-1}\right).
\label{eq:CIRC(U)}
\end{equation}
\normalsize
Note that there are only two none-zero $2 \times 2$ elements for any $k \geq3$.

As noted above, any $M \times M$ circulant matrix can be diagonalized with a commensurate Fourier matrix,
\begin{equation}
F^{M} = \left( F^{M}_{m,n}\right)  = \frac{1}{\sqrt{M}}\left(  e^{2 \pi i \frac{m n}{M}}\right) ,
\label{eq:FourM}
\end{equation}
with $m,n = 0,1,\ldots,M-1$. Correspondingly, the $2 \times 2$ block circulant matrix $U_{k}$ can be put in $2 \times 2$ block diagonal form using 
\begin{subequations}
\begin{align}
F&=F^{k} \otimes F^{2},
\label{eq:Four2k}\\
F U_{k} F^{\dagger} &= \begin{pmatrix}
U_{k,0} &  0  & \ldots & 0\\
0  &  U_{k,1} & \ldots & 0\\
\vdots & \vdots & \ddots & \vdots\\
0  &   0       &\ldots & U_{k,k-1}.
\label{eq:Four2k2}
\end{pmatrix}
\end{align}
\end{subequations}
For any $U_{k}$ the $l^{th}$ $2 \times 2$ block can be expressed as
\begin{equation}
\begin{split}
U_{k, l} = \frac{1}{2} \begin{bmatrix}
   \left(e^{-2\pi i \frac{l}{k}}e^{i\alpha}+e^{2\pi i \frac{l}{k}}e^{i\beta} \right) \sqrt{1-\rho} &    \left(-e^{-2\pi i \frac{l}{k}}e^{i\alpha}+e^{2\pi i \frac{l}{k}}e^{i\beta} \right) \sqrt{1-\rho} \\
{}+ \left(e^{-2\pi i \frac{l}{k}}-e^{2\pi i \frac{l}{k}}e^{i\left(\alpha+\beta \right) } \right) \sqrt{\rho} & {}+ \left(e^{-2\pi i \frac{l}{k}}+e^{2\pi i \frac{l}{k}}e^{i\left(\alpha+\beta \right) } \right) \sqrt{\rho} \\[6pt]
   \left(e^{-2\pi i \frac{l}{k}}e^{i\alpha}-e^{2\pi i \frac{l}{k}}e^{i\beta} \right) \sqrt{1-\rho} &     \left(-e^{-2\pi i \frac{l}{k}}e^{i\alpha}-e^{2\pi i \frac{l}{k}}e^{i\beta} \right) \sqrt{1-\rho} \\
{}+ \left(e^{-2\pi i \frac{l}{k}}+e^{2\pi i \frac{l}{k}}e^{i\left(\alpha+\beta \right) } \right) \sqrt{\rho} & {}+\left(e^{-2\pi i \frac{l}{k}}-e^{2\pi i \frac{l}{k}}e^{i\left(\alpha+\beta \right) } \right) \sqrt{\rho} \\
\end{bmatrix} \\
  l = 0,1,\ldots, k-1.
\end{split}
\label{eq:Block(U)}
\end{equation}
This is significant for determining the eigenvalues and eigenvectors of $U_{k}$.

\subsection{Powers of $U_{k}$}
Each step of the quantum walk on a cycle with $k$ nodes corresponds to an application of the quantum walk operator $U_{k}$. In terms of the initial state of the quantum walker $\left| \psi_{i} \rangle \right. $ the final state $\left| \psi_{f} \rangle \right.  $ after $N$ steps becomes
\begin{equation}
U_{k}^{N} \left| \psi_{i} \rangle \right. = \left| \psi_{f} \rangle \right. .
\label{eq:UN}
\end{equation}
By definition, \cite{Ivanova:1}  a cyclic group is a group that is generated by a single element, in the sense that every element of the cyclic group can be written as a power of a \textit{generator} element g in multiplicative notation. The set of elements of the finite cyclic group of order $n$, $G_{n}$, can then be written as
\begin{equation}
G_{n} = \left\lbrace 1=g^{0}=g^{n}, g, g^{2}, \ldots, g^{n-1}\right\rbrace 
\end{equation}
If there is no finite $n$ such that $g^{n} = 1$ then the cyclic group is infinite. Thus $U_{k}$ is the generator of a cyclic group. If the operator $U_{k}$ generates a finite group of order $N$ then the final state will equal the initial state in Eq. (\ref{eq:UN}).

It is well known that the set of solutions $\left\lbrace \lambda_{j}, \textbf{x}_{j} \right\rbrace$ satisfying the eigenvalue equation for an $M \times M$ matrix $A$ 
\begin{equation}
A \textbf{x}_{j} = \lambda_{j} \textbf{x}_{j}, \; j = 1, 2, \ldots, M
\end{equation}
also satisfy the relation
\begin{equation}
A^{n} \textbf{x}_{j} = \lambda^{n}_{j} \textbf{x}_{j}.
\end{equation}
Every $M \times M$ unitary matrix has a set of $M$ linearly independent (if not all orthogonal) eigenvectors $\left\lbrace \textbf{x}_{j} \right\rbrace $.  These eigenvectors form a basis for $\mathbb{C}_{M}$. An arbitrary initial state vector $ \left| \psi_{i} \rangle \right. $ in $\mathbb{C}_{2k}$ can then be expressed as a linear expansion of the eigenvectors of $U_{k}$
\begin{equation}
\left| \psi_{i} \rangle \right. = \sum_{j=1}^{2k} \alpha_{j}\textbf{x}_{j}
\end{equation}
so that
\begin{equation}
U_{k}^{N} \left| \psi_{i} \rangle \right. =  \sum_{j=1}^{2k} \alpha_{j} \lambda_{j}^{N} \textbf{x}_{j}
\end{equation}
Therefore, the condition $U_{k}^{N} \left| \psi_{i} \rangle \right. = \left| \psi_{i} \rangle \right.$ for an arbitrary initial state $\left| \psi_{i} \rangle \right.$ will be satisfied if and only if each of the eigenvalues of $U_{k}$ simultaneously satisfies
\begin{equation}
 \lambda_{j}^{N}  = 1.
 \label{eq:lambdaN}
\end{equation}
This directly leads to, 
\begin{equation}
U_{k}^{N} = I_{2k}.
\label{eq:UNk}
\end{equation}

\section{The eigenvalues of $U_{k}\left( \rho, \alpha, \beta\right) $}
The eigenvalues of any unitary matrix lie on the unit circle in the complex plane and take the general form $\lambda_{j} = e^{i \theta_{j}}$. The complete set of $2 k $ eigenvalues for any $U_{k}$ can be obtained in the union of the eigenvalues of each $2 \times 2$ block $U_{k, l}$ in Eq. (\ref{eq:Block(U)}). This yields
\begin{equation}
\begin{aligned} 
\lambda_{k,l}^{+} &= \frac{1}{2} e ^{-2\pi i \frac{l}{k}} \left( \left(1-e^{4\pi i \frac{l}{k}+i \delta} \right)\sqrt{\rho} \: + \: 2 \sqrt{e^{4\pi i \frac{l}{k}+i\delta} \left(1-\rho \sin^2 \left[ \frac{2\pi l}{k}+\frac{\delta}{2}\right]  \right) } \right) 
 \\
\lambda_{k,l}^{-} &=  \frac{1}{2} e ^{-2\pi i \frac{l}{k}} \left( \left(1-e^{4\pi i \frac{l}{k}+i \delta} \right)\sqrt{\rho}\: -\: 2 \sqrt{e^{4\pi i \frac{l}{k}+i\delta} \left(1-\rho \sin^2 \left[ \frac{2\pi l}{k}+\frac{\delta}{2}\right]  \right) } \right) 
 \\
\delta &= \alpha + \beta.
\end{aligned}
\label{eq:lambdakl}
\end{equation}
We note that the eigenvalue dependency on the $\alpha$ and $\beta$ parameters is exclusively $\delta = \alpha + \beta$.

Eqs. (\ref{eq:lambdaN}) and (\ref{eq:UNk}) would be satisfied when the eigenvalues simultaneously take the form of de Moivre numbers $\lambda_{j} = e^{2 \pi i \frac{m_{j}}{n_{j}}}$ where each pair $\left( m_{j}, n_{j}\right) $ are coprime, i.e., $\frac{m_{j}}{n_{j}}$ is a reduced rational, and $N = LCM\left( \left\lbrace n_{j} \right\rbrace  \right)$. With the appropriate branch cuts, the $\rho$ parameter which satisfies both of the equations
\begin{equation}
e^{-2\pi i \frac{m_{j}}{n_{j}}} \lambda_{k,l}^{\pm} = 1
\end{equation}
is
\begin{equation}
\rho_{l} = \frac{\sin^2\left(2\pi \frac{m_{j}}{n_{j}}-\frac{\delta}{2}  \right) }{\sin^2\left(2\pi \frac{l}{k}+\frac{\delta}{2}  \right) }=\frac{1-\cos \left( 4\pi \frac{m_{j}}{n_{j}}-\delta\right) }{1-\cos \left( 4\pi \frac{l}{k}+\delta\right) }.
\label{eq:rhol}
\end{equation}

\section{Solutions to $U_{k}^{N} = I_{2k} $}
Finding solutions $\left( N, \rho, \delta=\alpha + \beta\right)$ for $U_{k}^{N} = I_{2k} $} is essentially a matter of finding the $\left\lbrace \frac{m_{j}}{n_{j}}\right\rbrace$ and $\delta$ which make all available $\rho_{l}$ equal and $0 \leq \rho \leq 1$, then $N = \mathrm{LCM}\left( \left\lbrace n_{j}\right\rbrace  \right)$. In general the problem is ill-posed, for given $k$ there are $k$ equations with $k+2$ unknown parameters (this is considering $ \frac{m_{j}}{n_{j}}$ as a single parameter). The results in Eqs. (\ref{eq:lambdakl}) and (\ref{eq:rhol}) have two properties significant to this search. First we observe that in Eq. (\ref{eq:rhol}) for any odd integer $\eta$, $k=2\eta$ will produce the same $\left\lbrace \rho_{l} \right\rbrace $ as $k=\eta$ with double degeneracy \footnote{This does not mean that the eigenvalues of $U_{k}$ when $k=2\eta$ are doubly degenerate. Only that they simultaneously satisfy Eq. (\ref{eq:lambdaN}) with the same $\left( N, \rho, \delta = \alpha + \beta\right)$ as the $k=\eta$ case.} . This alone provides that $k=\eta$ and $k=2\eta$ have the same solution set. Second, a choice of $\delta$ which renders a $\rho_{l}$ as undefined will also render the corresponding eigenvalues in Eq. (\ref{eq:lambdakl}) as constants independent of $\rho$,
\begin{equation}
\begin{aligned} 
\lambda_{k,l}^{+} &=  e^{-2\pi i \frac{l}{k}} \\
\lambda_{k,l}^{-} &= - e^{-2\pi i \frac{l}{k}} .
\end{aligned}
\label{eq:lambdaklconst}
\end{equation}
Thus with choice of $\delta$ we can  reduce (sometimes significantly) the number of independent $\rho_{l}$ to be satisfied.

\subsection{$k \geq 2$ and $\rho = 0, 1$ case}
A particular set of solutions for all $k \geq 2$  occurs when $\rho = 0, 1$. When $\rho$ is either 0 or 1 only phase shifts and location shifts occur among the amplitudes. An initial arbitrary state will become periodic with successive applications of $U_{k}$ only with suitable values of $\alpha + \beta$. With appropriate  branch cuts, the eigenvalues in Eq. (\ref{eq:lambdakl}) become:

\begin{subequations}
\begin{align}
\mathrm{When}\: \rho &= 0 \nonumber \\
\lambda_{k, l} &= \left\lbrace e^{i \frac{\delta}{2}}, -e^{i \frac{\delta}{2}} \right\rbrace  \label{eq:lambdarho0}\\
\mathrm{When}\: \rho &= 1 \nonumber \\
\lambda_{k, l} &= \left\lbrace e^{-2\pi i \frac{l}{k}}, -e^{2\pi i \frac{l}{k}+\delta}\right\rbrace  \label{eq:lambdarho1}.
\end{align}
\end{subequations}

Eq. (\ref{eq:lambdarho0}) provides that for any $k \geq 2$, $U_{k}^{N} = I_{2k}$ when $\rho = 0$ and $\delta = 2\pi \frac{u}{v}$ where $\frac{u}{v}$ is a reduced rational and $N = 2v$. Eq. (\ref{eq:lambdarho1}) provides that for each $k \geq 2$, $U_{k}^{N} = I_{2k}$ when $\rho = 1$ and $\delta = 2\pi \frac{u}{v}$ and $N = \mathrm{LCM}\left[ 2, k, vk\right] $. Solutions for the $\rho = 0, 1$ case are summarized in Table \ref{table:rho=0,1}.
\begin{table} [ht]
\caption{Solutions $\left( N, \rho, \delta=\alpha + \beta\right)$ for $U_{k}^{N} = I_{2k} $ when $\rho = 0\:\mathrm{or}\: 1$ and $k \geq 2$.}
\label{table:rho=0,1}
\centering
\footnotesize\rm
\begin{tabular*}{\textwidth}{@{}l*{15}{@{\extracolsep{0pt plus12pt}}l}}
\br
\textbf{N} & $\rho$ & $\delta=\alpha + \beta$ \\
\br
$2v $&  $0$ & $2\pi \frac{u}{v}, \quad 0 <  \frac{u}{v} < 1$\\
\hline
$\mathrm{LCM}\left[ 2, k, vk\right] $ &  $1$ & $2\pi \frac{u}{v}, \quad 0 <  \frac{u}{v} < 1$\\
\br
\end{tabular*}
\end{table}

\subsection{$k = 2$ case}
The $k = 2$ case has $l = 0, 1$ and Eq. (\ref{eq:rhol}) yields 
\begin{equation}
\begin{aligned}
\rho_{0} = \rho_{1} = \frac{1-\cos \left( 4\pi \frac{m_{j}}{n_{j}}-\delta\right) }{1-\cos\left(\delta \right)}.
\end{aligned}
\label{eq:rho2}
\end{equation}
There is one function for $\rho$. It can be eliminated with $\delta = 0$, then Eq. (\ref{eq:lambdakl}) gives
\begin{equation}
\begin{aligned}
\lambda_{2,0}^{\pm} = \left\lbrace 1, -1\right\rbrace \\
\lambda_{2,1}^{\pm} = \left\lbrace -1, 1\right\rbrace
\end{aligned} 
\end{equation}
Thus when $\delta = 0$,  $U_{2}^{N} = I_{2k}$ is satisfied with any value for $\rho$ in the interval $0 < \rho < 1$ and $N=2$.

With a specified $\frac{m}{n}$ the range of  $\delta=2 \pi \frac{u}{v}$ maintaining $0 < \rho < 1$ in Eq. (\ref{eq:rho2}) is limited by
\begin{equation}
\begin{aligned}
\frac{ 2 m\left(\bmod{n} \right)}{2 n}< \frac{u}{v} < \frac{2 m\left(\bmod{n} \right) + n}{2 n}.
\end{aligned}
\label{eq:uvs}
\end{equation}
The only constraints we have for additional solutions are $0 < \left(\rho, \left\lbrace \frac{m_{j}}{n_{j}} \right\rbrace ,\frac{u}{v} \right) < 1$. Additional solutions can be found by proposing a ``seed" $\frac{m_{s}}{n_{s}}$, fixing a $\delta= 2\pi \frac{u}{v}$ within the limits in Eq. (\ref{eq:uvs}), finding the complete set of $\left\lbrace \frac{m_{j}}{n_{j}}\right\rbrace$ which hold $\rho$ constant, then $N = \mathrm{LCM}\left[ \left\lbrace n_{j}\right\rbrace \right] $.

\textbf{Example:} Starting with $\frac{m_{s}}{n_{s}} = \frac{2}{5}$ we obtain 
\begin{equation}
\begin{aligned}
\frac{2 }{5}< \frac{u}{v}< \frac{9}{10}.
\end{aligned}
\end{equation}
Within this interval we fix $\delta = 2\pi \frac{2}{3}$. Eq. (\ref{eq:rho2}) then gives
\begin{equation}
\begin{aligned}
 \rho = \frac{1-\cos \left( 4\pi \frac{2}{5}-2\pi \frac{2}{3}\right) }{1-\cos\left(2\pi \frac{2}{3} \right)} = \frac{2}{3}\left(1- \sin \left( \frac{7\pi}{30}\right) \right).
\end{aligned}
\end{equation}
The complete set of $\left\lbrace \frac{m_{j}}{n_{j}}\right\rbrace$ which keep $\rho$ constant is
\begin{equation}
\begin{aligned}
 \left\lbrace \frac{m_{j}}{n_{j}}\right\rbrace = \left\lbrace \frac{4}{15}, \frac{2}{5}, \frac{23}{30}, \frac{9}{10} \right\rbrace,
\end{aligned}
\end{equation}
and the resulting $N$ is
\begin{equation}
\begin{aligned}
N = \mathrm{LCM}\left[ \left\lbrace n_{j}\right\rbrace \right]  =  \mathrm{LCM}\left[ 15,5,30,10 \right] = 30.
\end{aligned}
\end{equation}

The solutions for $k=2$ are summarized in Table \ref{table:k=2}.
\begin{table} [ht]
\caption{Solutions $\left( N, \rho, \delta=\alpha + \beta\right)$ for $U_{k}^{N} = I_{2k} $ with $k=2$.}
\label{table:k=2}
\centering
\footnotesize\rm
\begin{tabular*}{\textwidth}{@{}l*{15}{@{\extracolsep{0pt plus12pt}}l}}
\br
\textbf{N} & $\rho$ & $\delta=\alpha + \beta$ \\
\br
$2$ & $0\leq\rho\leq1$ & 0 \\
\hline
$\mathrm{LCM}\left[\left\lbrace n_{j}\right\rbrace\right] $  & $\frac{1-\cos \left( 4\pi \frac{m_{j}}{n_{j}}-\delta\right) }{1-\cos\left(\delta \right)}$ & $2\pi \frac{u}{v}, \quad \mathrm{Seeded \: with \:} \frac{m_{s}}{n_{s}}$ \\ 
\br
\end{tabular*}
\end{table}

\subsection{$k = 3, 6$ cases}
As noted above, when $\eta$ is odd, $k=\eta$ and $k=2\eta$ will have equal solution sets. For $k = 3$ we have
\begin{equation}
\begin{aligned}
\rho_{0} &= \frac{1-\cos \left(4\pi \frac{m_{j}}{n_{j}} -\delta \right) }{1-\cos \left(\delta  \right) }\\
\rho_{1} &=  \frac{1-\cos \left(4\pi \frac{m_{j}}{n_{j}} -\delta \right) }{1-\cos \left(4\pi \frac{1}{3}+\delta  \right) }\\
\rho_{2} &=  \frac{1-\cos \left(4\pi \frac{m_{j}}{n_{j}} -\delta \right) }{1-\cos \left(4\pi \frac{2}{3}+\delta  \right) }.
\end{aligned}
\end{equation}
One of these can be eliminated with a choice of $\delta = \left( 0, 2\pi \frac{1}{3}, 2\pi \frac{2}{3}\right)$ and render constant valued $\lambda_{3,l}^{\pm}$ . At the same time, the other two $\rho_{l}\mathrm{'s}$ will be made equal:
\begin{subequations}
\begin{align}
\mathrm{When}\: \delta &= 0 \nonumber \\
\lambda_{3, 0} &= \left\lbrace 1, -1 \right\rbrace \nonumber \\
\rho_{1} = \rho_{2} &= \frac{2}{3} \left( 1 - \cos 4\pi \frac{m_{j}}{n_{j}} \right). \\
\mathrm{When}\: \delta &= 2\pi \frac{1}{3} \nonumber \\
\lambda_{3, 1} &= \left\lbrace -e^{2\pi i \frac{1}{6}}, e^{2\pi i \frac{1}{6}}\right\rbrace \nonumber \\
\rho_{0} = \rho_{2} &= \frac{2}{3} \left( 1- \cos \left( 4\pi \frac{m_{j}}{n_{j}} -  2\pi \frac{1}{3} \right) \right).  \\
\mathrm{When}\: \delta &= 2\pi \frac{2}{3} \nonumber \\
\lambda_{3, 2} &= \left\lbrace e^{2\pi i \frac{1}{3}}, -e^{2\pi i \frac{1}{3}}\right\rbrace \nonumber \\
\rho_{0} = \rho_{1} &= \frac{2}{3} \left( 1- \cos \left( 4\pi \frac{m_{j}}{n_{j}} -  2\pi \frac{2}{3} \right) \right). 
\end{align}
\end{subequations}
For each value of $\delta$, the $\left\lbrace \frac{m_{j}}{n_{j}} \right\rbrace$ which hold $\rho$ constant determine $N=\mathrm{LCM}\left[ \left\lbrace n_{j} \right\rbrace  \right] $. Solutions up to $N=30$ are summarized in Table \ref{table:k=3}. No solutions with other values of $0 \leq \delta \leq 2\pi$ are known.

\begin{table} [ht]
\caption{Solutions up to $N=30$ for $U_{k}^{N} = I_{2k} $ with $k=3$ and $k=6$.}
\centering
\footnotesize\rm
\label{table:k=3}
\begin{tabular*}{\textwidth}{@{}l*{15}{@{\extracolsep{0pt plus12pt}}l}}
\br
\textbf{N} & $\rho$ & $\delta= \alpha + \beta$ \\
\br
8 & $ \frac{2}{3}$ & 0 \\
\hline
10 & $\frac{5-\sqrt{5}}{6}$ & 0\\
\hline
12 & $\frac{1}{3}$ & 0, $\frac{2\pi}{3}$, $\frac{4\pi}{3}$\\
\hline
14 & $\frac{2}{3}\left(1-\cos \left(2 \pi \frac{1}{7} \right)  \right), \frac{2}{3}\left(1-\cos \left(2 \pi \frac{2}{7} \right)  \right) $ & 0\\
\hline
16 & $\frac{2-\sqrt{2}}{3}$ & 0\\
\hline
18 & $\frac{2}{3}\left(1-\cos \left(2 \pi \frac{1}{9} \right)  \right), \frac{2}{3}\left(1-\cos \left(2 \pi \frac{2}{9} \right)  \right) $  & 0, $\frac{2\pi}{3}$, $\frac{4\pi}{3}$\\
\hline
20 & $\frac{3-\sqrt{5}}{6}$, $\frac{3+\sqrt{5}}{6}$ & 0\\
\hline
22 & $\frac{2}{3}\left(1-\cos \left(2 \pi \frac{1}{11} \right)  \right), \frac{2}{3}\left(1-\cos \left(2 \pi \frac{2}{11} \right)  \right), \frac{2}{3}\left(1-\cos \left(2 \pi \frac{3}{11} \right)  \right) $ & 0\\
\hline
24 & $\frac{2-\sqrt{3}}{3}$ & 0, $\frac{2\pi}{3}$, $\frac{4\pi}{3}$ \\
\hline
24 & $\frac{2}{3}$ & $\frac{2\pi}{3}$, $\frac{4\pi}{3}$\\
\hline
26 & $\begin{pmatrix}
 \frac{2}{3}\left(1-\cos \left(2 \pi \frac{1}{13} \right)  \right), \frac{2}{3}\left(1-\cos \left(2 \pi \frac{2}{13} \right)  \right)\\
 \frac{2}{3}\left(1-\cos \left(2 \pi \frac{3}{13} \right)  \right), \frac{2}{3}\left(1-\cos \left(2 \pi \frac{4}{13} \right)  \right)
\end{pmatrix}$ & 0\\
\hline
28 & $\frac{2}{3}\left(1-\cos \left(2 \pi \frac{1}{14} \right)  \right), \frac{2}{3}\left(1-\cos \left(2 \pi \frac{3}{14} \right)  \right) $ & 0\\
\hline
30 & $\frac{7-\sqrt{5}-\sqrt{6 \left( 5-\sqrt{5} \right) }}{12} $, $\frac{7+\sqrt{5}-\sqrt{6 \left( 5+\sqrt{5} \right) }}{12} $, $ \frac{7-\sqrt{5}+\sqrt{6 \left( 5-\sqrt{5} \right) }}{12} $& 0, $\frac{2\pi}{3}$, $\frac{4\pi}{3}$\\
\hline
30 & $\frac{5-\sqrt{5}}{6}$ & $\frac{2\pi}{3}$, $\frac{4\pi}{3}$\\
\br
\end{tabular*}
\end{table}

\subsection{$k = 4$ case}
When $k=4$ we obtain
\begin{equation}
\begin{aligned}
\rho_{0} = \rho_{2} &=  \frac{1-\cos \left(4\pi \frac{m_{j}}{n_{j}} -\delta \right) }{1-\cos \left(\delta  \right) } \\
\rho_{1} = \rho_{3} &=  \frac{1-\cos \left(4\pi \frac{m_{j}}{n_{j}} -\delta \right) }{1+\cos \left(\delta  \right) }. 
\end{aligned}
\end{equation}

\begin{subequations}
\begin{align}
\mathrm{When}\: \delta &= 0 \nonumber \\
\lambda_{4, 0} &= \left\lbrace 1, -1 \right\rbrace, \quad \lambda_{4, 2} = \left\lbrace -1, 1 \right\rbrace \nonumber \\
\rho_{1} = \rho_{3} &= \frac{1}{2} \left( 1 - \cos 4\pi \frac{m_{j}}{n_{j}} \right). \\
\mathrm{When}\: \delta &= \pi \nonumber \\
\lambda_{4, 1} &= \left\lbrace -i, i \right\rbrace, \quad \lambda_{4, 3} = \left\lbrace i, -i \right\rbrace \nonumber \\
\rho_{0} = \rho_{2} &= \frac{1}{2} \left( 1+ \cos  4\pi \frac{m_{j}}{n_{j}} \right).
\end{align}
\end{subequations}
But we can also reduce the number of independent $\rho_{l}$ with $\delta = \frac{\pi}{2}, \frac{3\pi}{2}$;
\begin{subequations}
\begin{align}
\mathrm{When}\: \delta &= \frac{\pi}{2} \nonumber \\
\rho_{0} = \rho_{1} &= \rho_{2} = \rho_{3} = 1- \cos \left( 4\pi \frac{m_{j}}{n_{j}} -\frac{\pi}{2}\right). \\
\mathrm{When}\: \delta &= \frac{3\pi}{2} \nonumber \\
\rho_{0} = \rho_{1} &= \rho_{2} = \rho_{3} = 1- \cos \left( 4\pi \frac{m_{j}}{n_{j}} -\frac{3\pi}{2}\right). 
\end{align}
\end{subequations}

Solutions generated with each choice of $\delta$ up to $N=30$ are listed in Table \ref{table:k=4}. No solutions with other values of $0 \leq \delta \leq 2\pi$ are known.

\begin{table} [ht]
\caption{Solutions up to $N=30$ for $U_{k}^{N} = I_{2k} $ with $k=4$.}
\centering
\footnotesize\rm
\label{table:k=4}
\begin{tabular*}{\textwidth}{@{}l*{15}{@{\extracolsep{0pt plus12pt}}l}}
\br
$\textbf{N}$ & $\rho$ & $\delta=\alpha +\beta$ \\
\br
6 & $\frac{3}{4}$ & 0\\
\hline
8 & $\frac{1}{2}$ & 0, $\pi$\\
\hline
10 & $\frac{5-\sqrt{5}}{8}$, $\frac{5+\sqrt{5}}{8}$ & 0\\
\hline
12 & $\frac{1}{4}$ & 0, $\pi$\\
\hline
12 & $\frac{3}{4}$ & $\pi$\\
\hline
12 & $\frac{2-\sqrt{3}}{2}$ & $\frac{\pi}{2}$, $\frac{3\pi}{2}$\\
\hline
14 & $\frac{1}{2}\left(1-\sin\frac{3\pi}{14} \right), \frac{1}{2}\left(1+\sin\frac{\pi}{14} \right),\frac{1}{2}\left(1+\cos \frac{\pi}{7} \right)    $& 0\\
\hline
16 & $\frac{2-\sqrt{2}}{4}$, $\frac{2+\sqrt{2}}{4}$ & 0, $\pi$\\
\hline
16 & $\frac{2-\sqrt{2}}{2}$ & $\frac{\pi}{2}$, $\frac{3 \pi}{2}$\\
\hline
18 & $\frac{1}{2}\left(1-\cos\frac{2\pi}{9} \right),\frac{1}{2}\left(1-\sin \frac{\pi}{18} \right)  $ & 0\\
\hline
20 & $\frac{3-\sqrt{5}}{8}$, $\frac{3+\sqrt{5}}{8}$ & 0, $\pi$\\
\hline
20 & $\frac{5-\sqrt{5}}{8}$, $\frac{5+\sqrt{5}}{8}$ & $\pi$\\
\hline
20 & $\frac{4-\sqrt{10+2\sqrt{5}}}{4}$, $\frac{4-\sqrt{10-2\sqrt{5}}}{4}$ & $\frac{\pi}{2}, \frac{3 \pi}{2}$\\
\hline
22 & $\begin{pmatrix}
\frac{1}{2} \left( 1-\cos \frac{2\pi}{11} \right),\frac{1}{2}\left(  1-\sin \frac{3\pi}{22} \right),\frac{1}{2}\left( 1+\sin \frac{\pi}{22}\right) \\
 \frac{1}{2}\left(1+\sin  \frac{5\pi}{22}\right),\frac{1}{2}\left( 1+ \cos \frac{\pi}{11} \right) 
\end{pmatrix}$  & 0\\
\hline
24 & $\frac{2-\sqrt{3}}{4}$, $\frac{2+\sqrt{3}}{4}$ & 0, $\pi$\\
\hline
24 & $\frac{1}{2}$ & $\frac{\pi}{2}, \frac{3 \pi}{2} $\\
\hline
26 & $\begin{pmatrix}
\frac{1}{2} \left( 1-\cos \frac{2\pi}{13} \right),\frac{1}{2}\left(  1-\sin \frac{5\pi}{26} \right),\frac{1}{2}\left( 1-\sin \frac{\pi}{26}\right) \\
 \frac{1}{2}\left(1+\sin  \frac{3\pi}{26}\right),\frac{1}{2}\left( 1+ \cos \frac{3\pi}{13}\right) ,\frac{1}{2}\left( 1+\cos \frac{\pi}{13} \right) 
\end{pmatrix}$ & 0\\
\hline
28 & $ \frac{1}{2}\left(1-\cos \frac{\pi}{7} \right),\frac{1}{2}\left( 1-\sin \frac{\pi}{14}\right),\frac{1}{2}\left(1+\sin \frac{3\pi}{14} \right) $& 0, $\pi$\\
\hline
28 & $ \frac{1}{2}\left(1+\cos \frac{\pi}{7} \right),\frac{1}{2}\left( 1+\sin \frac{\pi}{14}\right),\frac{1}{2}\left(1-\sin \frac{3\pi}{14} \right) $ & $\pi$\\
\hline
28 & $ 1-\sin\frac{\pi}{7}, 1-\cos \frac{\pi}{14},1-\cos \frac{3\pi}{14}$ & $\frac{\pi}{2}, \frac{3 \pi}{2}$\\
\hline
30 & $\frac{7-\sqrt{5}-\sqrt{6 \left( 5-\sqrt{5} \right) }}{16} $, $\frac{7+\sqrt{5}-\sqrt{6 \left( 5+\sqrt{5} \right) }}{16}$, $ \frac{7-\sqrt{5}+\sqrt{6 \left( 5-\sqrt{5} \right) }}{16} $,  $\frac{7+\sqrt{5}+\sqrt{6 \left( 5+\sqrt{5} \right) }}{16} $& 0\\
\br
\end{tabular*}
\end{table}
\subsection{$k = 5, 10$ and $k = 8$ cases}
The $k = 5, 10$ and $k=8$ cases can be reduced to a minimum of two independent forms for the $\rho_{l}$ with a choice in $\delta$.

When $k=5, 10$ and $\delta = \left( 0, 2\pi \frac{1}{5}, 2\pi \frac{2}{5}, 2\pi \frac{3}{5}, 2\pi \frac{4}{5}\right) $ we obtain:
\begin{equation}
\begin{aligned}
\rho' &=  \frac{4}{5+\sqrt{5}} \left( 1- \cos \left( 4\pi \frac{m'_{j}}{n'_{j}} -\delta\right) \right) \\
\rho'' &= \frac{4}{5-\sqrt{5}} \left( 1- \cos \left( 4\pi \frac{m''_{j}}{n''_{j}} -\delta\right) \right) . 
\end{aligned}
\end{equation}
Only a finite combination of $\left\lbrace \frac{m'_{j}}{n'_{j}},  \frac{m''_{j}}{n''_{j}}\right\rbrace$ are known that satisfy $\rho' = \rho''$ for each $\delta$. 

When $k=8$ and $\delta = \left( 0, 2\pi \frac{1}{4}, 2\pi \frac{2}{4}, 2\pi \frac{3}{4} \right)$ we obtain:
\begin{equation}
\begin{aligned}
\rho' &=  1- \cos \left( 4\pi \frac{m'_{j}}{n'_{j}} -\delta\right)  \\
\rho'' &= \frac{1}{2} \left( 1- \cos \left( 4\pi \frac{m''_{j}}{n''_{j}} -\delta\right) \right),
\end{aligned}
\end{equation}
and again only a finite combination of $\left\lbrace \frac{m'_{j}}{n'_{j}},  \frac{m''_{j}}{n''_{j}}\right\rbrace$ are known to satisfy $\rho' = \rho''$ for each $\delta$.

The known solutions for $k=5,10$ and $k=8$ are listed in Table  \ref{table:k=5,10,8}.

\def\arraystretch{1.3}%
\begin{table} [ht]
\caption{Solutions for $U_{k}^{N} = I_{2k} $ with $k=5,10$ and $k=8$.}
\centering
\footnotesize\rm
\label{table:k=5,10,8}
\begin{tabular*}{\textwidth}{@{}l*{15}{@{\extracolsep{0pt plus12pt}}l}}%
%\begin{tabular}{c c c c c}
\br
\multicolumn{5}{c}{\textbf{k=5,10}} \\
\br
\textbf{N} & $\rho$ & $\delta=\alpha +\beta $ & $\left\lbrace \frac{m'_{j}}{n'_{j}}\right\rbrace $ &  $\left\lbrace \frac{m''_{j}}{n''_{j}}\right\rbrace $ \\
\hline
\multicolumn{1}{ l }{\multirow{10}{*}{60} } &
\multicolumn{1}{ c }{\multirow{5}{*}{$\frac{5-\sqrt{5}}{10}$} } &
\multicolumn{1}{ l } {0} & {$\left\lbrace \frac{1}{12}, \frac{5}{12}, \frac{7}{12}, \frac{11}{12}  \right\rbrace $} & {$\left\lbrace \frac{1}{20}, \frac{9}{20}, \frac{11}{20}, \frac{19}{20}  \right\rbrace $} \\ 
\cline{3-5}
\multicolumn{1}{ c }{} & \multicolumn{1}{ c }{} &
\multicolumn{1}{ l } {$\frac{2\pi}{5}$} & {$\left\lbrace \frac{1}{60}, \frac{11}{60}, \frac{31}{60}, \frac{41}{60}  \right\rbrace $} & {$\left\lbrace \frac{1}{20}, \frac{3}{20}, \frac{11}{20}, \frac{13}{20}  \right\rbrace $}  \\ 
\cline{3-5}
\multicolumn{1}{ c }{} & \multicolumn{1}{ c }{} &
\multicolumn{1}{ l }{$\frac{4\pi}{5}$} & {$\left\lbrace \frac{7}{60}, \frac{17}{60}, \frac{37}{60}, \frac{47}{60}  \right\rbrace $} & {$\left\lbrace \frac{3}{20}, \frac{1}{4}, \frac{13}{20}, \frac{3}{4}  \right\rbrace $} \\ 
\cline{3-5}
\multicolumn{1}{ c }{} & \multicolumn{1}{ c }{} &
\multicolumn{1}{ l }{$\frac{6\pi}{5}$} & {$\left\lbrace \frac{13}{60}, \frac{23}{60}, \frac{43}{60}, \frac{53}{60}  \right\rbrace $} & {$\left\lbrace \frac{1}{4}, \frac{7}{20}, \frac{3}{4}, \frac{17}{20}  \right\rbrace $}  \\ 
\cline{3-5}
\multicolumn{1}{ c }{} & \multicolumn{1}{ c }{} &
\multicolumn{1}{ l }{$\frac{8\pi}{5}$} & {$\left\lbrace \frac{19}{60}, \frac{29}{60}, \frac{49}{60}, \frac{59}{60}  \right\rbrace $} & {$\left\lbrace \frac{7}{20}, \frac{9}{20}, \frac{17}{20}, \frac{19}{20}  \right\rbrace $}  \\ 
\cline{2-5}
\multicolumn{1}{ c }{} & \multicolumn{1}{ c }{\multirow{5}{*}{$\frac{5+\sqrt{5}}{10}$} } &
\multicolumn{1}{ l } {0} & {$\left\lbrace \frac{3}{20}, \frac{7}{20}, \frac{13}{20}, \frac{17}{20}  \right\rbrace $} & {$\left\lbrace \frac{1}{12}, \frac{5}{12}, \frac{7}{12}, \frac{11}{12}  \right\rbrace $} \\ 
\cline{3-5}
\multicolumn{1}{ c }{} & \multicolumn{1}{ c }{} &
\multicolumn{1}{ l } {$\frac{2\pi}{5}$} & {$\left\lbrace \frac{1}{4}, \frac{9}{20}, \frac{3}{4}, \frac{19}{20}  \right\rbrace $} & {$\left\lbrace \frac{1}{60}, \frac{11}{60}, \frac{31}{60}, \frac{41}{60}  \right\rbrace $}  \\ 
\cline{3-5}
\multicolumn{1}{ c }{} & \multicolumn{1}{ c }{} &
\multicolumn{1}{ l }{$\frac{4\pi}{5}$} & {$\left\lbrace \frac{1}{20}, \frac{7}{20}, \frac{11}{20}, \frac{17}{20}  \right\rbrace $} & {$\left\lbrace \frac{7}{60}, \frac{17}{60}, \frac{37}{60}, \frac{47}{60}  \right\rbrace $} \\ 
\cline{3-5}
\multicolumn{1}{ c }{} & \multicolumn{1}{ c }{} &
\multicolumn{1}{ l }{$\frac{6\pi}{5}$} & {$\left\lbrace \frac{3}{20}, \frac{9}{20}, \frac{13}{20}, \frac{19}{20}  \right\rbrace $} & {$\left\lbrace \frac{13}{60}, \frac{23}{60}, \frac{43}{60}, \frac{53}{60}  \right\rbrace $}  \\ 
\cline{3-5}
\multicolumn{1}{ c }{} & \multicolumn{1}{ c }{} &
\multicolumn{1}{ l }{$\frac{8\pi}{5}$} & {$\left\lbrace \frac{1}{20}, \frac{1}{4}, \frac{11}{20}, \frac{3}{4}  \right\rbrace $} & {$\left\lbrace \frac{19}{60}, \frac{29}{60}, \frac{49}{60}, \frac{59}{60}  \right\rbrace $}  \\ 
\br
\multicolumn{5}{c}{\textbf{k=8}} \\
\br
\textbf{N} & $\rho$ & $\delta=\alpha +\beta $ & $\left\lbrace \frac{m'_{j}}{n'_{j}}\right\rbrace $ &  $\left\lbrace \frac{m''_{j}}{n''_{j}}\right\rbrace $\\
\hline
\multicolumn{1}{ l }{\multirow{4}{*}{24} } &
\multicolumn{1}{ c }{\multirow{4}{*}{$\frac{1}{2}$} } &
\multicolumn{1}{ l } {0} & {$\left\lbrace \frac{1}{12}, \frac{5}{12}, \frac{7}{12}, \frac{11}{12}  \right\rbrace $} & {$\left\lbrace \frac{1}{8}, \frac{3}{8}, \frac{5}{8}, \frac{7}{8}  \right\rbrace $} \\ 
\cline{3-5}
\multicolumn{1}{ l }{} & \multicolumn{1}{ l }{} &
\multicolumn{1}{ l } {$\frac{\pi}{2}$} & {$\left\lbrace \frac{1}{24}, \frac{5}{24}, \frac{13}{24}, \frac{17}{24}  \right\rbrace $} & {$\left\lbrace \frac{1}{4}, \frac{1}{2}, \frac{3}{4}  \right\rbrace $}  \\ 
\cline{3-5}
\multicolumn{1}{ l }{} & \multicolumn{1}{ l }{} &
\multicolumn{1}{ l }{$\pi$} & {$\left\lbrace \frac{1}{6}, \frac{1}{3}, \frac{2}{3}, \frac{5}{6}  \right\rbrace $} & {$\left\lbrace \frac{1}{8}, \frac{3}{8}, \frac{5}{8}, \frac{7}{8}  \right\rbrace $} \\ 
\cline{3-5}
\multicolumn{1}{ l }{} & \multicolumn{1}{ l }{} &
\multicolumn{1}{ l }{$\frac{3\pi}{2}$} & {$\left\lbrace \frac{7}{24}, \frac{11}{24}, \frac{19}{24}, \frac{23}{24}  \right\rbrace $} & {$\left\lbrace \frac{1}{4}, \frac{1}{2}, \frac{3}{4}  \right\rbrace $}  \\ 
\br
\end{tabular*}
\end{table}
\def\arraystretch{1.0}%

\subsection{$k = 7, 9$ and $k \geq 11$ cases}
All remaining $U_{k}$ operators have no known exact solutions for $U_{k}^{N}=I_{2k}$ other than the particular solutions when $\rho = 0, 1$.

Approximate solutions with arbitrarily small error can always be constructed by choosing values for $0 < \rho <1$ and $0 \leq \delta \leq 2\pi$  and for each $\rho_{l}$ find the set of rational $\left\lbrace \frac{m_{j}}{n_{j}}\right\rbrace$ which satisfy $\rho_{l}=\rho \pm \epsilon$ with $\epsilon$ as small as desired, then $N=\mathrm{LCM}\left[ \left\lbrace  n_{j}\right\rbrace \right]$.

\section{Revivals of special states}
Satisfying  $U_{k}^{N} |\psi_{i}\rangle = |\psi_{i}\rangle$ without $U_{k}^{N} = I_{2k}$ puts special constraints on $|\psi_{i}\rangle$. The initial state vector must be expressible as a linear expansion over a subset of the eigenvectors of $U_{k}$ which have simultaneous ``de Moivre" eigenvalues;
\begin{equation}
\begin{aligned}
|\psi_{i}\rangle &= \sum_{\lbrace \mathbf{x'}_{j}\rbrace \subset \lbrace \mathbf{x}_{j}\rbrace} \alpha'_{j} \mathbf{x'}_{j} \\
U^{N}_{k}|\psi_{i}\rangle &= \sum_{\lbrace \mathbf{x'}_{j}\rbrace \subset \lbrace \mathbf{x}_{j}\rbrace} \alpha'_{j} \lambda'^{N}_{j} \mathbf{x'}_{j}\\
\mathrm{Such \: that} & \mathrm{ \: each \: \lambda'_{\textit{j}} \: satisfies}\\
\lambda'^{N}_{j} &= 1. 
\end{aligned}
\label{equ:spcstate}
\end{equation}

\textbf{Example:} Derive a $k=4$ state for which $N=5$. The $\rho_{l}$ to be satisfied are:

\begin{equation}
\begin{aligned}
\rho_{0} &= \frac{1-\cos \left( 4\pi \frac{m_{j}}{n_{j}}-\delta\right) }{1-\cos \left( \delta\right) }\\
\rho_{1} &= \frac{1-\cos \left( 4\pi \frac{m_{j}}{n_{j}}-\delta\right) }{1-\cos \left( 4\pi \frac{1}{4}+\delta\right) }\\
\rho_{2} &= \frac{1-\cos \left( 4\pi \frac{m_{j}}{n_{j}}-\delta\right) }{1-\cos \left( 4\pi \frac{2}{4}+\delta\right) }\\
\rho_{3} &= \frac{1-\cos \left( 4\pi \frac{m_{j}}{n_{j}}-\delta\right) }{1-\cos \left( 4\pi \frac{3}{4}+\delta\right) }.
\end{aligned}
\end{equation}

\begin{figure}[h!]
\includegraphics[scale=0.38]{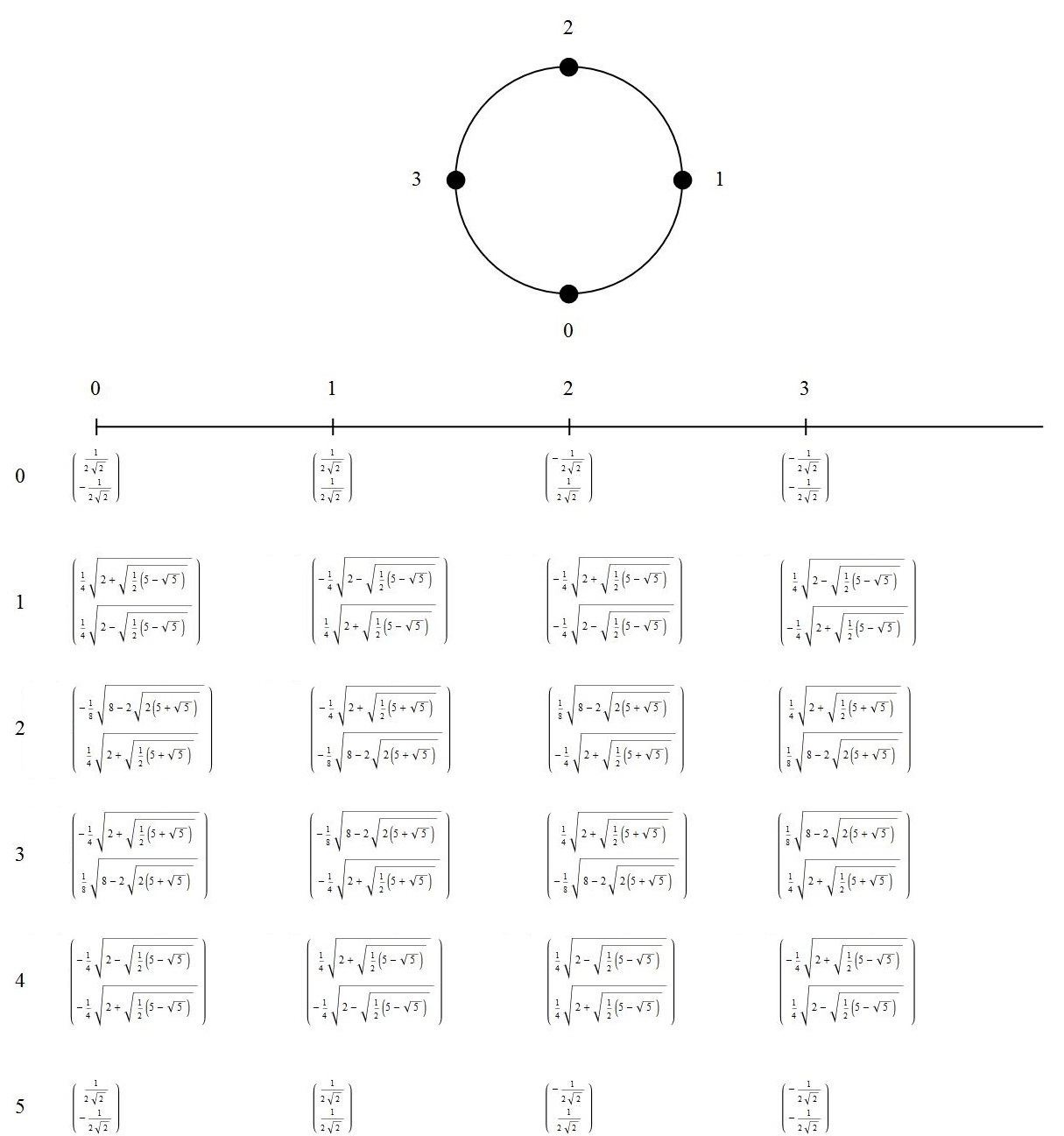}
\caption{A $k = 4$ quantum state with a period of $N = 5$, 
 $U_{4}\left(\rho = \frac{5-\sqrt{5}}{8}, \alpha = 0, \beta = 0 \right) $.}
\label{fig:k4N5}
\end{figure}

We choose $\delta = 0$ (specifically $\alpha=0, \beta=0$) and this eliminates $\rho_{0}$ and $\rho_{2}$. The corresponding eigenvalues for $l=0,2$ in Eq. (\ref{eq:lambdakl}) are 
\begin{equation}
\begin{aligned}
\lambda^{+}_{4,0} &= 1\\
\lambda^{-}_{4,0} &= -1\\
\lambda^{+}_{4,2} &= -1\\
\lambda^{-}_{4,2} &= 1.
\end{aligned}
\end{equation}
The remaining $\rho_{1}$ and $\rho_{3}$ must be made equal such that the corresponding eigenvalues are fifth roots of 1. The possibilities are listed in Table \ref{table:k=4,N=5}. 
\begin{center}
\begin{table}[ht]
\caption{Solutions for $k=4, \frac{m}{n}=\left\lbrace \frac{1}{5}, \frac{2}{5}, \frac{3}{5}, \frac{4}{6} \right\rbrace , \alpha=0, \beta=0$.}
\centering 
\label{table:k=4,N=5}
\begin{tabular}{|c|c c|c c|c c|c c|}
\hline
 \multirow{2}[2]{*}{$\rho_{l}$} & \multirow{2}[2]{*}{$\frac{m}{n}=\frac{1}{5}$}
& $\lambda^{+}_{4,l}$
&  \multirow{2}[2]{*}{$\frac{m}{n}=\frac{2}{5}$}
& $\lambda^{+}_{4,l}$ & \multirow{2}[2]{*}{$\frac{m}{n}=\frac{3}{5}$}
& $\lambda^{+}_{4,l}$ & \multirow{2}[2]{*}{$\frac{m}{n}=\frac{4}{5}$}
& $\lambda^{+}_{4,l}$ \\ \cline{3-3} \cline{5-5} \cline{7-7} \cline{9-9}
 & &  $\lambda^{-}_{4,l}$ & & $\lambda^{-}_{4,l}$ & & $\lambda^{-}_{4,l}$ & & $\lambda^{-}_{4,l}$  \\ \hline

 \multirow{2}[2]{*}{$\rho_{1}$} & \multirow{2}[2]{*}{$\frac{5+\sqrt{5}}{8}$}
& $e^{-i \frac{2\pi}{5}}$
&  \multirow{2}[2]{*}{$\frac{5-\sqrt{5}}{8}$}
& $e^{-i \frac{\pi}{5}}$ & \multirow{2}[2]{*}{$\frac{5-\sqrt{5}}{8}$}
& $e^{-i \frac{\pi}{5}}$ & \multirow{2}[2]{*}{$\frac{5+\sqrt{5}}{8}$}
& $e^{-i \frac{2\pi}{5}}$ \bigstrut \\ \cline{3-3} \cline{5-5} \cline{7-7} \cline{9-9}
 & &  $e^{-i \frac{3\pi}{5}}$ & & $e^{-i \frac{4\pi}{5}}$ & & $e^{-i \frac{4\pi}{5}}$ & & $e^{-i \frac{3\pi}{5}}$ \bigstrut \\ \hline

 \multirow{2}[2]{*}{$\rho_{3}$} & \multirow{2}[2]{*}{$\frac{5+\sqrt{5}}{8}$}
& $e^{i \frac{3\pi}{5}}$
&  \multirow{2}[2]{*}{$\frac{5-\sqrt{5}}{8}$}
& $e^{i \frac{4\pi}{5}}$ & \multirow{2}[2]{*}{$\frac{5-\sqrt{5}}{8}$}
& $e^{i \frac{4\pi}{5}}$ & \multirow{2}[2]{*}{$\frac{5+\sqrt{5}}{8}$}
& $e^{i \frac{3\pi}{5}}$ \bigstrut \\ \cline{3-3} \cline{5-5} \cline{7-7} \cline{9-9}
 & &  $e^{i \frac{2\pi}{5}}$ & & $e^{i \frac{\pi}{5}}$ & & $e^{i \frac{\pi}{5}}$ & & $e^{i \frac{2\pi}{5}}$ \bigstrut \\ \hline
 \end{tabular}
\end{table}
\end{center}
We see that with the chosen conditions there are two possible values, $\rho =\frac{5\pm\sqrt{5}}{8}$. We will choose $\rho =\frac{5 - \sqrt{5}}{8}$ which corresponds with $\frac{m_{l}}{n_{l}}=\left\lbrace \frac{2}{5}, \frac{3}{5}\right\rbrace $. The eigenvalues satisfying $ \lambda'^{5}_{j} = 1$ are $\lambda'_{j}=\left\lbrace \lambda^{+}_{4,0}, \lambda^{-}_{4,1}, \lambda^{-}_{4,2}, \lambda^{+}_{4,3} \right\rbrace$.  The corresponding eigenvectors of $U_{k}$ can be obtained from the eigenvectors of the $2\times2$ diagonal blocks of $U_{k,l}$ in Eq. (\ref{eq:Block(U)}) using the Fourier matrix of Eq. (\ref{eq:Four2k}). The eigenvectors corresponding with $\lambda^{-}_{4,1}, \lambda^{+}_{4,3}$ are combined to yield the initial state in Fig. \ref{fig:k4N5}.

\section{Conclusion}
The diagonal blocks and eigenvalues given by Eqs. (\ref{eq:Block(U)}) and (\ref{eq:lambdakl}) capture the essential structure and eigenvalue spectrum of the cyclic operator $U_{k}$. On inspection, the $l,k$ and $\delta$  dependence shows that when $k$ is odd valued $U_{k}$ and $U_{2k}$ can have equal eigenvalue spectrums, $U_{2k}$ being doubly degenerate. The condition that the eigenvalue spectrum of the $l^{th}$ $2\times2$ block consists of de Moivre numbers of the form $e^{2\pi i \frac{m}{n}}$ is that $\rho_{l}=\frac{1-\cos \left( 4\pi \frac{m_{j}}{n_{j}}-\delta\right) }{1-\cos\left( 4\pi \frac{l}{k}+\delta\right)}$ where $ \frac{m_{j}}{n_{j}}$ is relatively prime and $\delta=\alpha+\beta$, Eq. (\ref{eq:rhol}). For a given $k$, anytime values for  $ \frac{m_{j}}{n_{j}}$ and $\delta$ can be found to make all $\rho_{l}$ equal and $0<\rho_{l}<1$ a solution to $U_{k}^{N}=I_{2k}$ is obtained. As noted above the problem is ill posed with more unknown parameters than independent equations. However, for each $k$, values for $\delta$ can be found which reduce the number of independent $\rho_{l}$ to a minimum. If reduced to a single functional form for all $\rho_{l}$ there will be an infinite number of solutions. This is the case for $k=\left\lbrace 2,3,4,6\right\rbrace $. For $k=\left\lbrace 5,8,10\right\rbrace$ the minimum number of independent $\rho_{l}$ can be reduced to two. Consider the $k=8$ case, when $\delta=0$ the eight possible $\rho_{l}$ simplify to two independent forms. We then want to solve 
\begin{equation}
\rho=1-\cos \left( 4\pi \frac{m'}{n'}\right) = \frac{1}{2}\left( 1-\cos \left( 4\pi \frac{m''}{n''}\right) \right).
\label{eq:rho8}
\end{equation}
Inspection gives $\frac{m'}{n'}=\left\lbrace  \frac{1}{12},\frac{5}{12},\frac{7}{12},\frac{11}{12}\right\rbrace$ and $\frac{m''}{n''}=\left\lbrace  \frac{1}{8},\frac{3}{8},\frac{5}{8},\frac{7}{8}\right\rbrace$ with $\rho=\frac{1}{2}$ and $N=24$ in all cases. Yet there is no proof that any other rational values could or could not satisfy Eq. (\ref{eq:rho8}). A similar situation is found for $k=\left\lbrace 5, 10\right\rbrace$.

For all other values of $k$ the various $\rho_{l}$ are reducible to no fewer than three or more independent forms and have no known solutions. This leaves it open to question whether some $U_{k}^{N}=I_{2k}$ have any solutions or even an infinite number of solutions.

Our interest in solutions for $U_{k}^{N}=I_{2k}$ has been purely academic. The course has revealed some interesting properties and structures of the cyclic operator  $U_{k}$ which may prove useful in development of quantum walk search algorithms.

\section{Acknowledgments}
I wish to thank the anonymous reviewers for their help in pointing out areas in which the original manuscript could be made clearer. I wish to also thank Zolt\'{a}n Dar\'{a}zs, Dardo Goyeneche, Salvador E. Venegas-Andraca, and Yutaka Shikano for reading the manuscript and pointing out additional relevant references.

\section*{References}
\bibliographystyle{unsrt}
\bibliography{QWrefs}

\end{document}